\documentclass{IEEEcsmag}

\usepackage[colorlinks,urlcolor=blue,linkcolor=blue,citecolor=blue]{hyperref}
\expandafter\def\expandafter\UrlBreaks\expandafter{\UrlBreaks\do\/\do\*\do\-\do\~\do\'\do\"\do\-}
\usepackage{upmath,color}
\usepackage{makecell}
\usepackage[normalem]{ulem}
\usepackage{xcolor}

\newif\ifshowrevisions
\showrevisionsfalse

\ifshowrevisions
    \newcommand{\revadd}[1]{\textcolor{blue}{#1}}
    \newcommand{\revdel}[1]{\textcolor{red}{\sout{#1}}}
\else
    \newcommand{\revadd}[1]{#1}
    \newcommand{\revdel}[1]{}
\fi

\jvol{XX}
\jnum{XX}
\paper{8}
\jmonth{Month}
\jname{Publication Name}
\jtitle{Publication Title}
\pubyear{2021}

\begin{document}

\sptitle{Department: Visualization Viewpoints}

\title{Envisioning Mobile Data Visualization Libraries for Digital Health}

\author{Bongshin Lee}
\affil{Yonsei University, Seoul, Republic of Korea}

\author{Seongjae Bae}
\affil{Yonsei University, Seoul, Republic of Korea}

\author{Mengying Li}
\affil{University of Maryland, College Park, MD, USA}

\author{Eun Kyoung Choe}
\affil{University of Maryland, College Park, MD, USA}

\markboth{DEPARTMENT}{DEPARTMENT}

\begin{abstract}\looseness-1Mobile health (mHealth) applications support health management through \revadd{the collection and visualization of rich data}, yet the quality of the visualizations varies widely. A key limitation %is %the suboptimal design of visualizations for small-screen devices.
\revadd{lies in the challenge of effectively visualizing temporally dense, irregular, and context-dependent health data within the constrained mobile interfaces.} %, yet current developer tools provide little support for such health-specific semantics.} 
We argue that this gap is partly driven by a lack of specialized developer tools. Existing libraries primarily target desktop or general-purpose mobile use, providing limited support for health-specific semantics such as normal ranges, thresholds, and goals. As a result, developers often resort to custom solutions that are inconsistent or hard to interpret. We therefore advocate for dedicated mobile visualization libraries tailored to personal health data and mobile contexts, and discuss key design considerations including intelligent defaults, built-in health annotations, and fluid interaction. Such libraries can lower the barrier \revadd{to producing effective visualizations and make mHealth data easier for users to interpret.}
\end{abstract}

\maketitle

%In the header at the top of page 1, please indicate whether your article is a Theme Article, Feature Article, or Department submission. If it is a Theme Article, include the special issue title as the description. If it is a Feature Article, please provide a 3-4 word phrase reflecting the topic of the article. If it is a Department submission, please name the\break department.\vadjust{\pagebreak} 

% [R2-1][R2-2][R3-1] Explains why health data are hard to visualize, why interpretive support is needed, and how missing health-specific context can make charts ambiguous.
% [R1-3][R2-3] Adds external references, including Dunn et al. and Chan et al., to reduce reliance on self-citations and support health data context and mobile health visualization claims.
\chapteri{M}obile technologies have become central to how people monitor and understand their health. From fitness trackers and smartwatches to continuous glucose monitors and diet-logging apps, these tools generate continuous streams of personal health and behavior data. As of 2024, more than 350,000 mobile health (mHealth) applications are available globally, reflecting the rapid growth of digital tools for personal health management.\footnote{mHealth Apps Statistics and Facts (2026), https://market.biz/mhealth-apps-statistics/} 
Most mHealth apps employ data visualization to transform raw data into visual representations that support understanding, self-reflection, and increased awareness, helping individuals achieve their tracking goals.

Yet despite their widespread adoption, \revadd{research on data visualization in mHealth apps is still in its infancy~\cite{chan2025shape}, and creating effective and accessible data visualizations for mobile health remains challenging. 
%Mobile contexts introduce unique constraints, including limited screen space and interaction capabilities, while also requiring representations that are easily interpretable by lay individuals\revdel{ [1]}%~\cite{lee2020reaching}}}.
Personal health data are inherently multidimensional, encompassing a wide range of measures such as heart rate, blood pressure, stress, sleep, physical activity, weight, and medication. These data vary not only in their units and scales but also in their temporal characteristics. Some are collected continuously (e.g., heart rate), while others are recorded intermittently or as discrete events (e.g., weight and medication intake). Interpreting these health data involves various observations, including relative vs. absolute change, missing data, thresholds and outliers, and frequency~\cite{dunn2025typology}. Moreover, meaningful interpretation depends on contextual information, such as relevant events (e.g., meals and exercise), appropriate reference ranges (e.g., target glucose range), and personal baselines.}

\revadd{Because mHealth apps are used by people with diverse levels of health and visualization literacy, they need to provide interpretive support that helps users determine whether observed patterns are expected, concerning, changing over time, or uncertain. Furthermore, mHealth visualizations need to communicate this complex and context-dependent information within the constraints of small screens.}
Designing such visualizations draws on knowledge of visualization principles, which may not always be readily available within typical mobile development workflows. 
%\revadd{Mobile app developers lacking visualization knowledge may produce misleading visualization, for example, a stress chart that connects non-consecutive measurement dates (Figure~\ref{fig:stressTrend}).}
%Many mobile app developers are not visualization experts and often lack foundational visualization knowledge needed to design representations that are both effective and appropriate for small-screen, everyday use by lay individuals.

\begin{figure*} [t]
    \centering
    \begin{tabular}{cp{0.2cm}cp{0.2cm}c}
    \includegraphics[width=.27\linewidth]{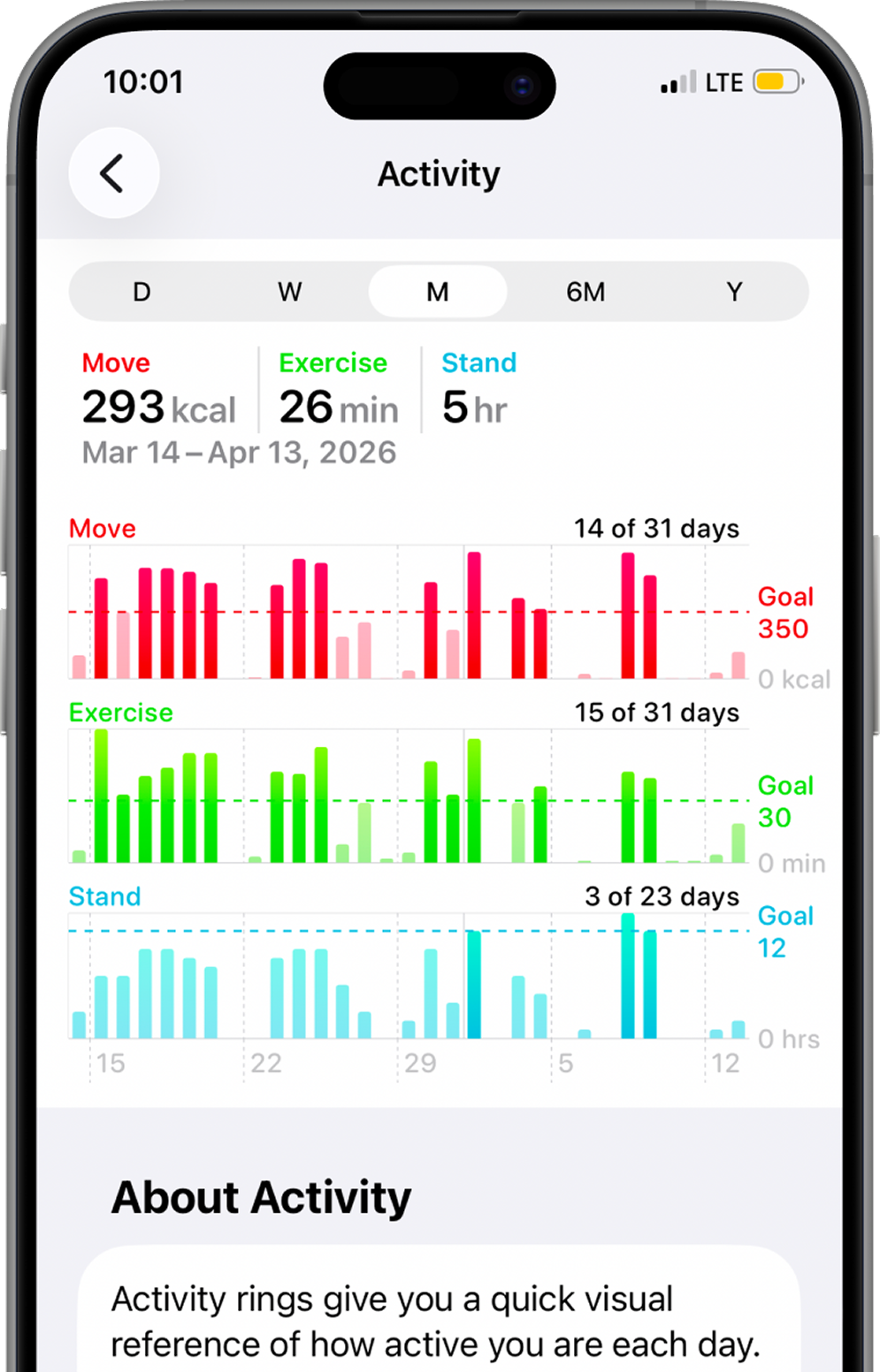} & &
    \includegraphics[width=.27\linewidth]{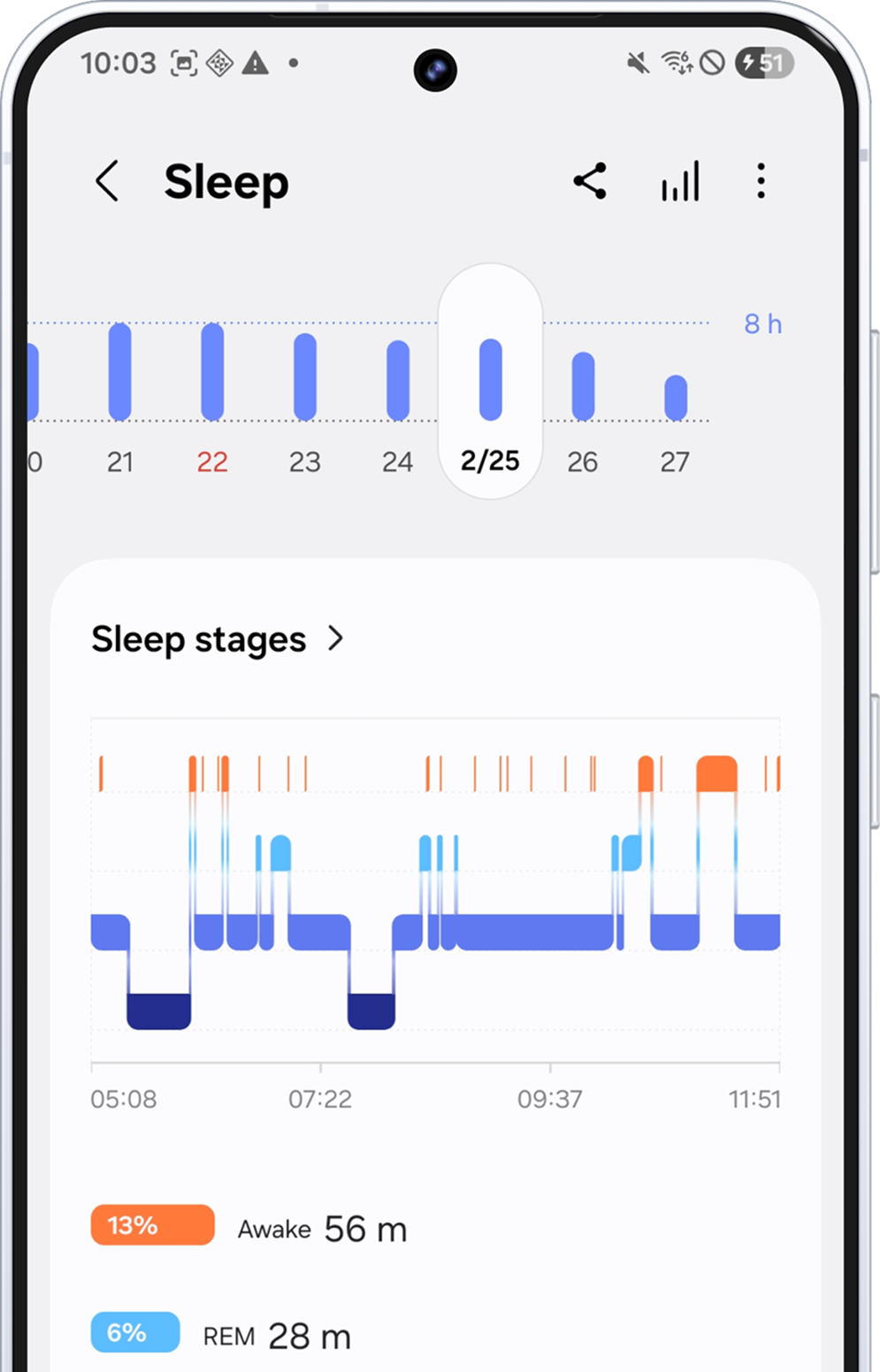} & &
    \includegraphics[width=.27\linewidth]{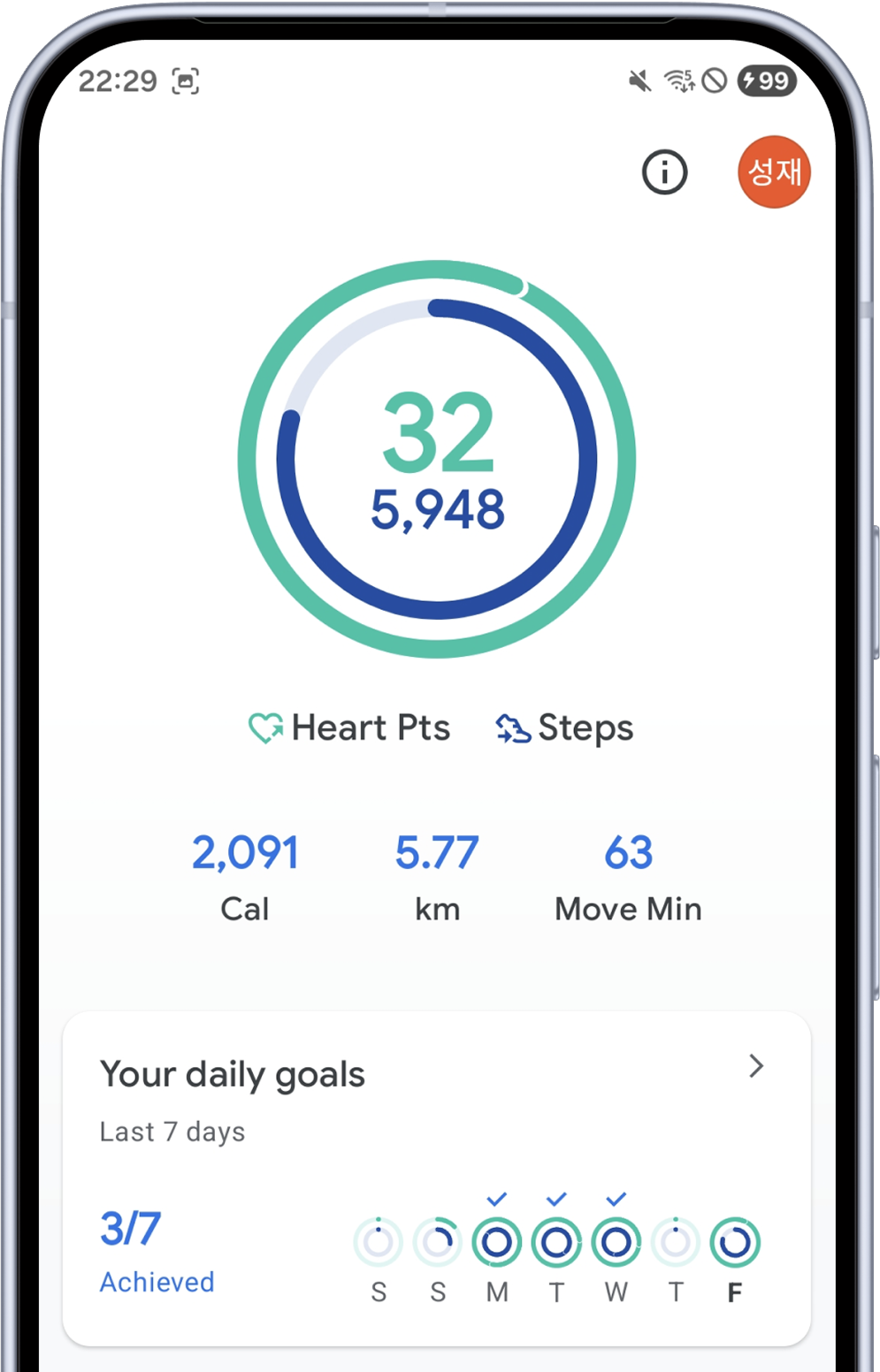} \\
    {\footnotesize Apple Health} & & {\footnotesize Samsung Health} & & {\footnotesize Google Fit}
    \end{tabular}
    \caption{Health data visualizations from three major mobile health platforms: the Activity view in Apple Health (left), the Sleep view in Samsung Health (middle), and the Home page in Google Fit (right).}
    \label{fig:mHealthVis}
\end{figure*}

%[R3-3]
This mismatch between the growing demand for mHealth visualizations and the limited availability of specialized design resources points to a clear gap in our visualization ecosystem. While desktop and web-based visualization libraries such as D3~\cite{bostock2011d3} and Vega-Lite~\cite{satyanarayan2016vega} have advanced how people author and share visualizations, mobile visualization cannot be adequately achieved by simply resizing desktop visualizations to fit smaller screens~\cite{lee2021mobile}. 

Existing libraries designed for mobile devices (e.g., Victory Native,\footnote{https://nearform.com/open-source/victory-native/} React Native SVG Charts,\footnote{https://github.com/JesperLekland/react-native-svg-charts} MPAndroidChart,\footnote{https://github.com/PhilJay/MPAndroidChart} \revadd{fl\_chart,\footnote{https://pub.dev/packages/fl\_chart} and Syncfusion Flutter Charts\footnote{https://pub.dev/packages/syncfusion\_flutter\_charts}}) provide charting capabilities. 
However, these libraries are developed for general-purpose or cross-platform use rather than mHealth applications. 
They provide little support for encoding health-relevant semantics such as normal ranges, thresholds, and goals. %, app developers must shoulder the burden of ensuring effective, meaningful, and interpretable visual representations of health data.
\revadd{In mHealth apps, technically correct charts may still fail to convey meaningful insights when health-specific context is missing. For example, a glucose chart without target ranges may obscure sustained periods of hypo- or hyperglycemia, making it difficult for users to assess whether their glucose levels are within a healthy range. %Similarly, a stress chart that connects non-consecutive measurement dates may misrepresent the underlying trend (Figure~\ref{fig:stressTrend}).} 
Because existing charting tools and libraries do not account for this kind of health-specific context,} app developers must shoulder the burden of ensuring effective, meaningful, and interpretable visual representations of health data.

We call for dedicated mobile data visualization libraries for mHealth apps that are tailored to the unique characteristics of personal health data and mobile contexts, \revadd{and provide reusable support for common health visualization needs}. 
Drawing on prior research, including our own work in personal health informatics and mobile data visualization, we identify key design considerations: \revadd{intelligent defaults, built-in annotation capabilities, fluid interaction, time as a core dimension, support for diverse display contexts, inclusivity and accessibility, and configurable disclosure of sensitive health data.}

%[1-5]
\revdel{These libraries should go beyond providing reusable charting components to also embed visualization principles grounded in health data interpretation and mobile interaction. By lowering both technical and design barriers, they can accelerate the development of mHealth apps and promote more consistent and interpretable data visualization practices across the mHealth ecosystem. Ultimately, they contribute to making personal health data more comprehensible, actionable, and inclusive.}

%\caption{Note that ``Figure'' is spelled out. There is a period after the figure number, followed by one space. It is good practice to briefly explain the significance of the figure in the caption. (From [``Title''],$^1$ used with permission.)}\vspace*{-5pt}

\begin{figure*}
    \centering
    \begin{tabular}{cccc}
        \includegraphics[width=.225\linewidth]{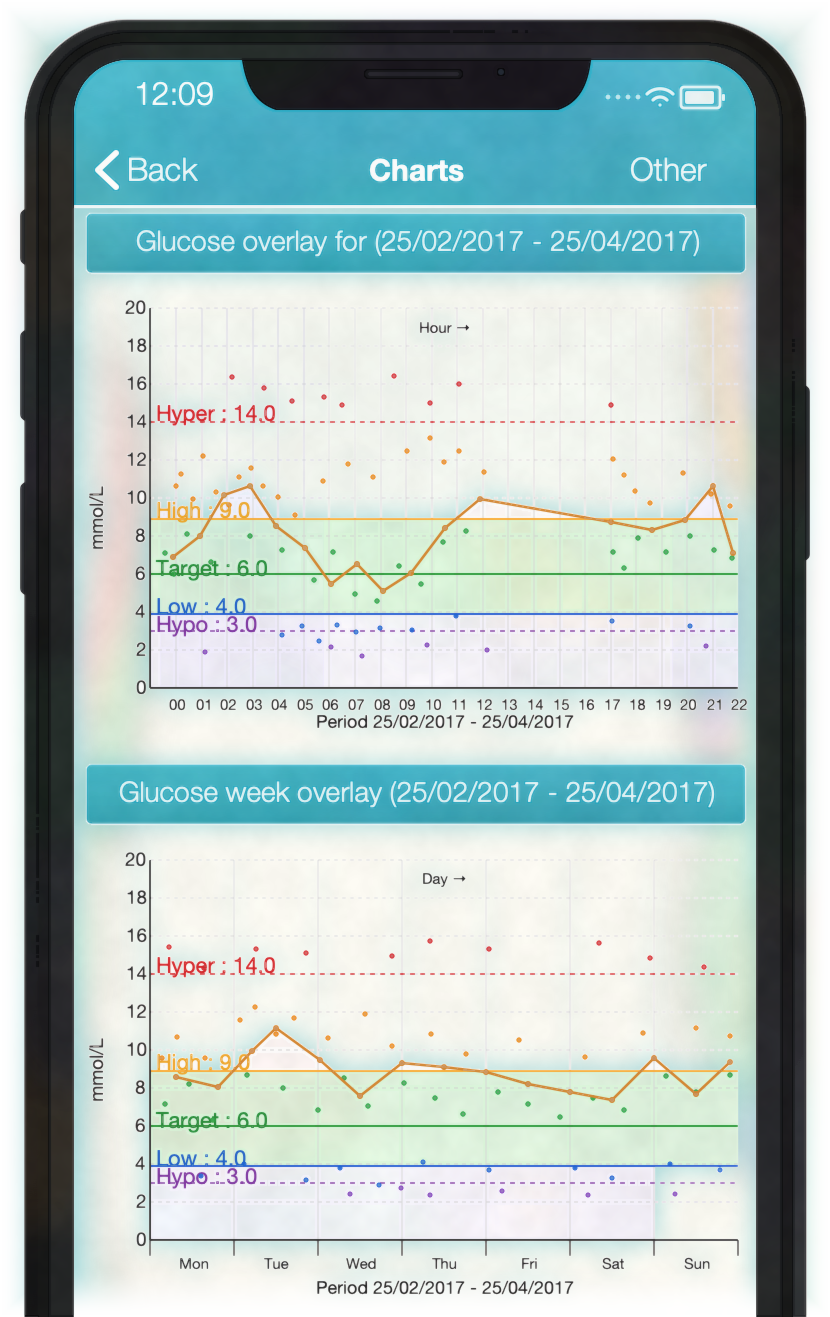} &
        \includegraphics[width=.225\linewidth]{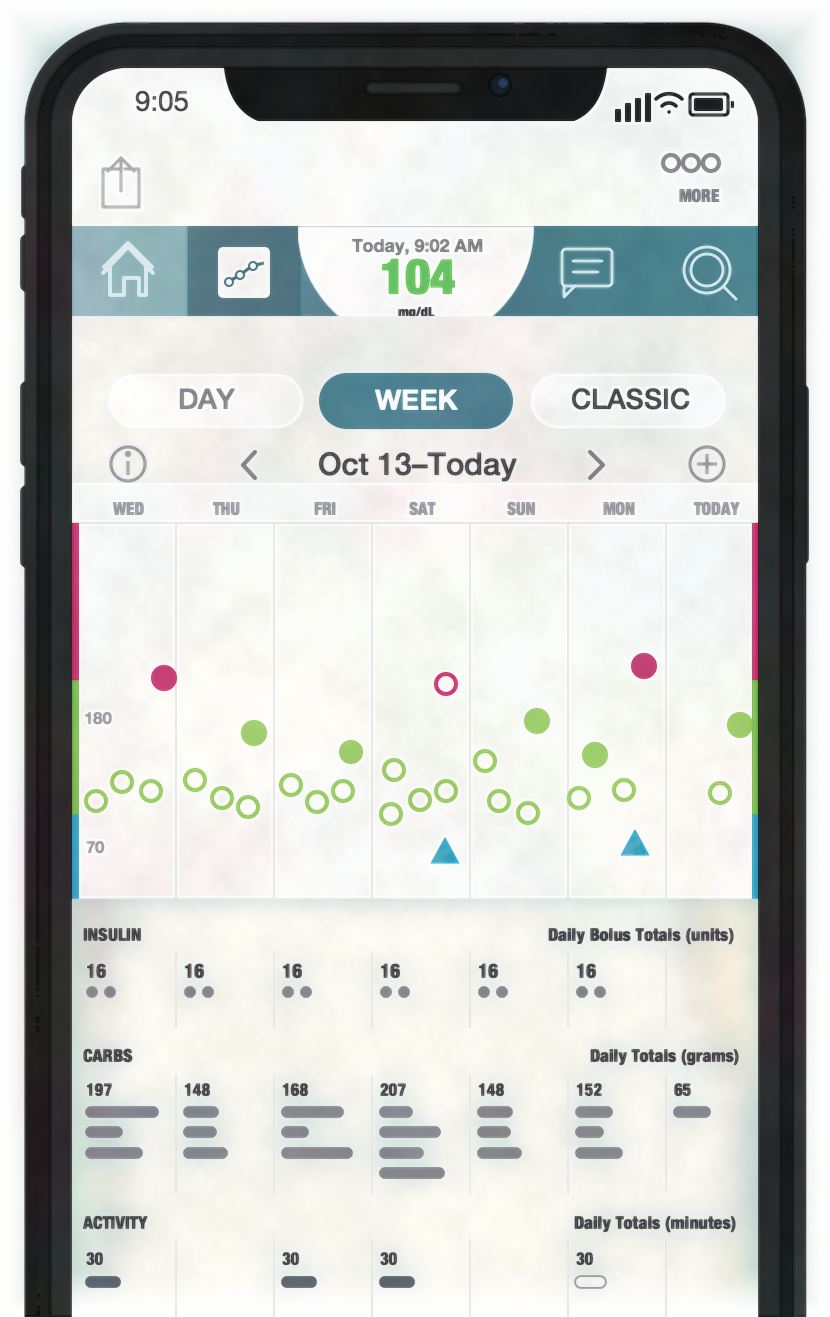} &
        \includegraphics[width=.225\linewidth]{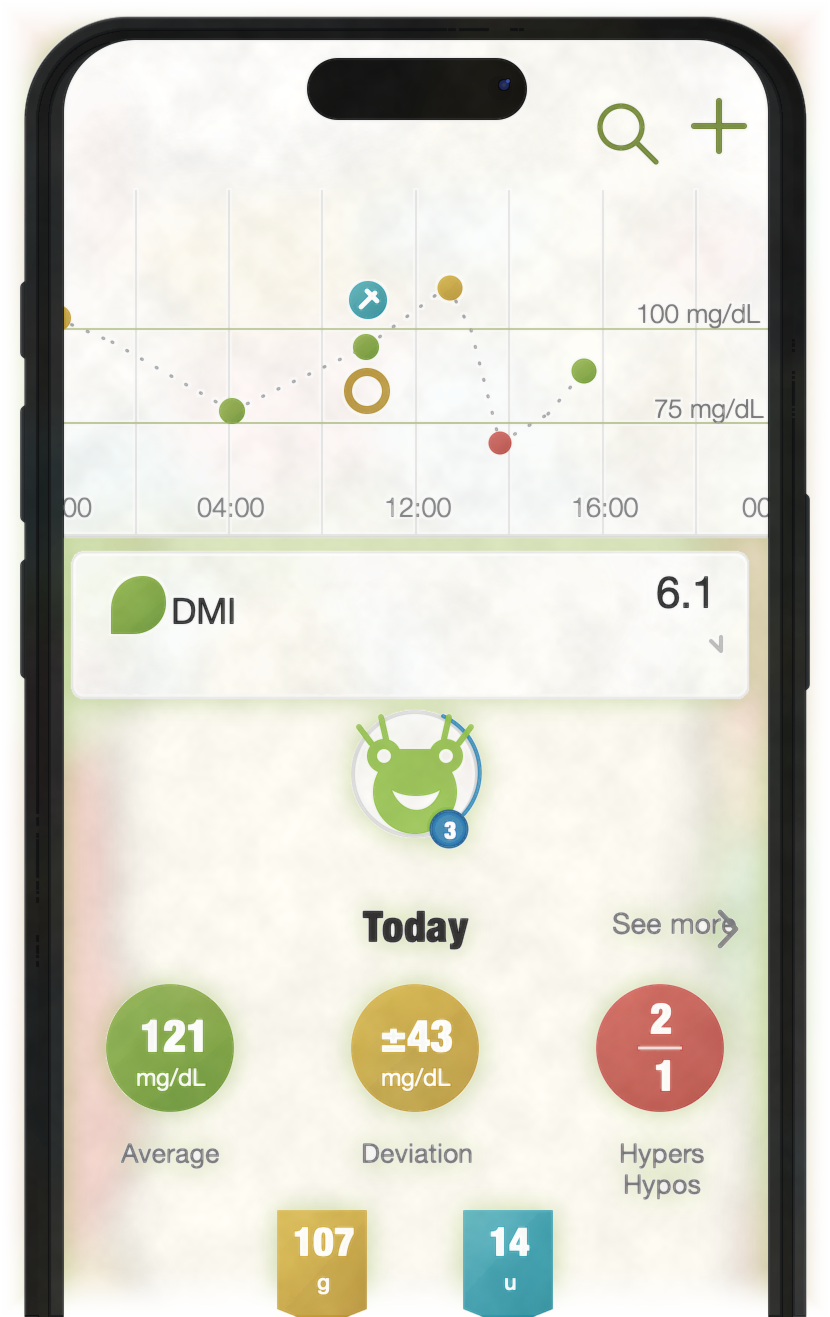} &
        \includegraphics[width=.225\linewidth]{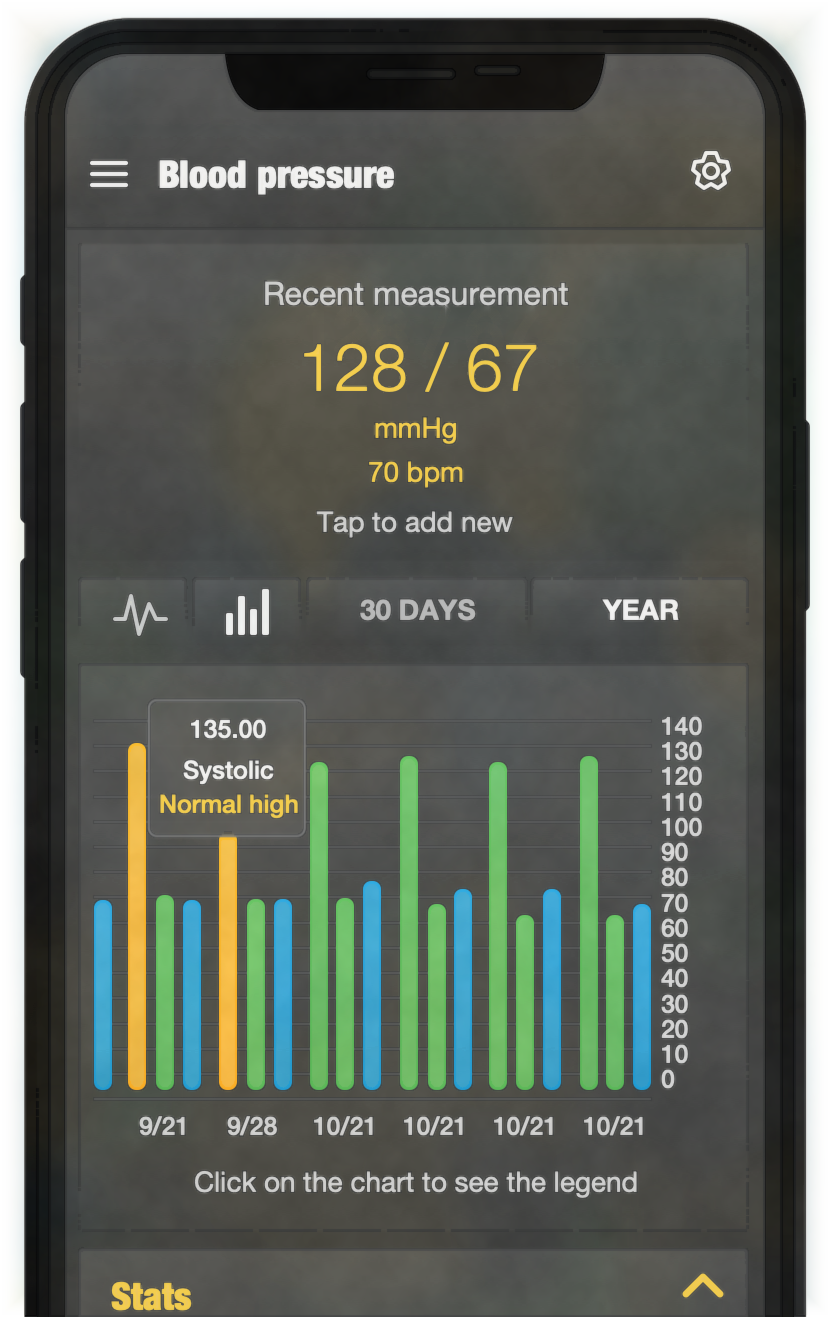} \\
        {\footnotesize Long-term trend view} & {\footnotesize Weekly view} & {\footnotesize Daily summary} & {\footnotesize Blood pressure} 
    \end{tabular}
    \caption{Examples of mHealth visualizations in diabetes management apps that illustrate issues such as visual clutter, missing legends, ambiguous encodings, and inappropriate chart types. These examples are created using ChatGPT 5.5 based on publicly available screenshots. %are overly complex or insufficiently specified, making them difficult to understand. As these examples were reproduced using ChatGPT 5.4 based on screenshots available on the web, certain details may not be accurate.
    %\revadd{AI-synthesized representative examples of mHealth visualizations in diabetes management apps, created using ChatGPT 5.4 based on publicly available screenshots selected by the authors. The examples illustrate recurring issues such as visual clutter, ambiguous encodings, and insufficient contextual cues without identifying or singling out specific applications.}
    }
    \label{fig:poorVis}
\end{figure*}

\section{MOBILE HEALTH VISUALIZATIONS}

Prior research has demonstrated the power of combining visual representations with effective interaction to support self-reflection by making abstract data more concrete and interpretable~\cite{choe2017understanding}. %Well-designed visualizations enable users to identify trends, detect anomalies, and reflect on their behaviors over time. 
In addition to research prototypes featuring creative health data visualizations, such as UbiFit Garden~\cite{consolvo2008activity} and Fish’n’Steps~\cite{lin2006fish}, data visualizations are now widely used in mHealth apps to support people’s understanding of their health data~\cite{chan2025shape}. Modern mHealth apps, including widely used commercial ones (e.g., MyFitnessPal, Noom, and Strava), commonly rely on charts and summary views to communicate activities, diet, and progress toward personal goals over time. Platform-level apps such as Apple Health, Samsung Health, and Google Fit employ various visualizations to present diverse health metrics (e.g., resting heart rate and sleep) and activities (e.g., step count and exercise) collected from multiple devices, sensors, and services (Figure \ref{fig:mHealthVis}). 

Despite their widespread adoption, effective visualization of health data on mobile devices remains challenging. Health data are inherently heterogeneous in type, scale, and temporal resolution, often combining continuous sensor streams, intermittent self-reports, and event-based records. When coupled with the limited screen real estate, imprecise touch input, and fragmented attention characteristic of mobile use, these properties make it difficult to present health information in ways that are both clear and meaningful. 

Prior work on mobile data visualization has emphasized that simply adapting desktop visualizations to mobile devices is insufficient, as mobile contexts require rethinking visual encodings and interaction techniques to support brief, on-the-go use and reduced attention~\cite{lee2021mobile}.
These challenges are further compounded by the fact that many mobile app developers may have limited exposure to the visualization principles needed to design effective representations for small-screen, everyday use by lay individuals. 

As a result, many mHealth visualizations remain poorly suited to small-screen smartphones. Common issues include overly dense charts, excessive reliance on color encoding, truncated axes, and the direct adaptation of desktop visualizations to mobile contexts~\cite{alshehhi2022analysis}. These shortcomings lead to visual clutter, limited readability and understandability, and ambiguous encodings, making it difficult for users to accurately and effectively interpret their health data (Figure~\ref{fig:poorVis} illustrates several such examples from diabetes management apps). 
To address this, we need to provide better design support for creating visualizations that are effective, interpretable, and mobile-appropriate.

\section{DATA VISUALIZATION LIBRARIES}

% [R3-3][R3-4] Expands representative native and cross-platform libraries, including Flutter libraries, and clarifies limits of general-purpose charting primitives.
Web-based visualization libraries have significantly advanced how people author, customize, and share visualizations by facilitating the creation of systems with data visualizations. 
However, mobile data visualization presents different challenges that cannot be addressed by simply resizing desktop visualizations to fit smaller screens~\cite{lee2021mobile}. Mobile devices impose constraints on screen real estate, interaction precision, and user attention. In the context of health data, these constraints are further compounded by challenges related to the heterogeneity of health data~\cite{meyer2016visualization}, as well as by high temporal density, the need for contextual interpretation, and personal nature of the data.

Several visualization libraries have been developed specifically for mobile or cross-platform environments, including Victory Native, React Native SVG Charts, MPAndroidChart, \revadd{fl\_chart, and Syncfusion Flutter Charts}. These libraries provide basic charting primitives and enable developers to render common visualization types on mobile devices. 
For example, Victory was originally built for the web as an ecosystem of composable React components for building interactive data visualizations, leveraging the power of D3 and SVG (Scalable Vector Graphics) for rendering. It was later abstracted so that custom React elements could be used to render the various parts of the data visualizations. React Native SVG Charts leverage SVG to create data visualization in React Native applications. MPAndroidChart provides a set of charts specifically tailored to the Android platform. \revadd{Similarly, fl\_chart and Syncfusion Flutter Charts provide customizable chart components for Flutter, %applications, extending this general-purpose charting support to 
another major cross-platform mobile development environment.} 

Apple’s Swift Charts\footnote{https://developer.apple.com/documentation/Charts} is a declarative charting framework introduced as part of SwiftUI to support data visualization across Apple platforms, including iOS, iPadOS, watchOS, and macOS. Designed with mobile contexts in mind, Swift Charts enables developers to create common chart types, such as line, bar, area, and point charts using a concise, declarative syntax that integrates seamlessly with SwiftUI’s layout, animation, and state management model. In mobile contexts, Swift Charts emphasizes responsiveness, adaptability, and platform consistency. Charts automatically adapt to different screen sizes, orientations, and accessibility settings, and integrate with system-level features such as dark mode and VoiceOver. The framework also supports basic interactive behaviors, including selection, highlighting, and animated transitions, which are well suited to touch-based interaction on small screens.

However, these libraries for mobile are primarily designed for general-purpose data visualization or visual consistency across platforms, rather than for health-specific applications. 
%As a result, they offer limited support for encoding health-relevant semantics such as clinically meaningful ranges, personalized thresholds, goals, or contextual events. 
\revadd{Their limitation is not an inability to render health-related charts, but rather the lack of built-in support for health-specific semantics. Developers still need to decide how to aggregate dense temporal data, represent thresholds, goals, and baselines, and connect contextual events such as medication or symptoms with numerical trends. Recurring mHealth chart types, such as sleep-stage charts and activity-goal summaries, often need to be approximated using generic marks or implemented through custom drawing. As a result, developers often build an additional health-specific visualization layer on top of general-purpose charting primitives.}

\revdel{Moreover, these libraries typically focus on visual rendering rather than interpretive support, leaving critical design decisions—such as how to aggregate data, highlight deviations from normal ranges, or convey uncertainty and variability—to individual developers. This places a substantial burden on individual app developers to ensure that visualizations are not only visually appealing but also effective, meaningful, and interpretable for end users. In addition, mobile visualization calls for new approaches to interaction design and may require supporting lightweight and transient exploration \revdel{[11]} with fewer and simpler manipulations for making changes to data, representation, presentation, and view parameters} \revdel{[1].} 

In summary, a gap remains between the capabilities of current general-purpose mobile visualization libraries and the requirements for developing mHealth applications that effectively support understanding, reflection, and informed health management.

\section{DESIGN CONSIDERATIONS FOR MOBILE HEALTH VISUALIZATION LIBRARIES}
To guide the development of visualization libraries for mHealth apps, we outline key design considerations informed by prior research on mobile data visualization and the unique characteristics of mobile contexts.

\subsection{DC1. Appropriate Default Representations}
% [R1-1][R2-4][R3-5] Makes intelligent defaults more specific to recurring mHealth data structures and chart types.

Visualization libraries should provide appropriate default representations that enable mobile app developers without visualization expertise to easily create visualizations for helping end users make sense of their health data. Well-chosen defaults would play a critical role in ensuring clarity and effectiveness. 

They should be grounded in perceptual principles and tailored to the characteristics of personal health data through appropriate visual encodings, data aggregation, and clear visual hierarchies. They should also consider common mHealth contexts, such as brief, frequent interactions and users' varying levels of data and health literacy. 
\revadd{Default representations could support recurring mHealth data structures. For example, activity data could default to compact progress-to-goal views, glucose data to configurable target-range bands, and sleep data to stage-based timelines. These defaults should remain interpretable while allowing customization for different health conditions, user goals, and application styles.}

By embedding these design considerations into default representations, visualization libraries can help ensure a baseline level of quality and consistency, enabling even those without visualization expertise to produce health visualizations that are usable, interpretable, and aligned with established best practices. Ultimately, this helps improve users’ ability to accurately understand their data and supports more informed reflection and health management.

\subsection{DC2. Annotations for Contextual Information and Health Semantics}
mHealth visualization libraries should offer built-in support for annotations that convey contextual information and health-specific semantics. In mHealth settings, raw numerical values or trends are often difficult for users to interpret without additional context, particularly for users with limited health or data literacy. Annotations can help address this challenge by highlighting clinically or behaviorally meaningful events (e.g., medication intake, symptom onset, or missed activities), calling attention to notable changes or trends, and explaining why these patterns may matter. Embedding such contextual cues directly in the visualization helps transform abstract data into information that is more personally relevant and actionable.

Effective annotation support can go beyond static labels to include semantics that reflect users' personal conditions, goals, and contexts. This includes indicating what constitutes a “healthy,” “normal,” or “concerning” range, adapting explanations to individual goals or health conditions. By integrating annotation mechanisms directly into visualization libraries, developers can more easily create views that support interpretation, reflection, and sensemaking. Such support enables the development of mHealth visualizations that help users understand their health status and make informed decisions based on it.

\subsection{DC3. Fluid Interaction for Navigation and Exploration}
% [R2-5][R3-6] Clarifies fluid interaction and adds touch-based detail-on-demand and scrubbing examples.
%To support effective navigation and exploration on small screens, visualization libraries should provide fundamental interaction techniques such as zooming, filtering, temporal navigation, and comparison.
\revadd{While generally useful in desktop environments as well, fluid interaction is particularly important on mobile phones}, where screen real estate is limited and interactions are often brief and situational. 
\revadd{By fluid interaction, we mean enabling users to easily access details and move seamlessly between different views and levels of detail using simple, familiar gestures.} %\revadd{Building on Rey et al.’s notion of databiting, we define “fluid interaction” as lightweight, transient interactions that support brief inspection from compact mobile overviews without requiring a separate analysis context\revadd{~\cite{rey2024databiting}}.}\revdel{Well-designed interactions can help users focus on relevant subsets of data, examine changes over time, and compare values or periods of interest, all while maintaining context.} 
\revadd{To support effective navigation and exploration of health data on mobile phones, visualization libraries should provide fundamental interaction techniques including selection, details-on-demand, zooming, and temporal navigation.
For example, for timelines showing heart rate, libraries could provide built-in support for displaying a tooltip or selecting an item on tap, zooming in or out by pinching, moving to the previous or next period while panning horizontally, and showing details while scrubbing (panning after pressing and holding).
This could be achieved through configurable mappings from basic gestures, including tap, pinch, swipe, and drag, to common visualization operations.}
Treating fluid interaction capabilities as first-class features in visualization libraries can help ensure that common exploratory tasks are supported consistently across mHealth apps.

These interactions should be designed for mobile contexts and be extensible to support multimodal input, including touch and voice. \revadd{For example, as demonstrated by Data@Hand~\cite{kim2021data}, speech-based queries and time manipulation, combined with touch, can facilitate filtering and comparison of health data on smartphones.} Beyond basic input, multimodal support can enable more flexible interaction styles, allowing users to choose the most convenient or accessible modality depending on their context and needs~\cite{lee2021post}. Leveraging such modalities can lower interaction barriers, reduce cognitive and physical effort, and enable more natural engagement with health data, particularly in everyday and on-the-go settings.

\subsection{DC4. Treating Time as a First-Class Citizen}
Personal health data are inherently temporal, reflecting patterns, routines, and changes that unfold extended periods. As such, mHealth visualization libraries should treat time as a first-class design dimension rather than a secondary attribute. This entails native support for temporal structures and interactions that go beyond simple linear timelines, enabling users to explore data across multiple time scales while maintaining context.

\begin{figure}
    \centering
    \includegraphics[width=0.85\linewidth]{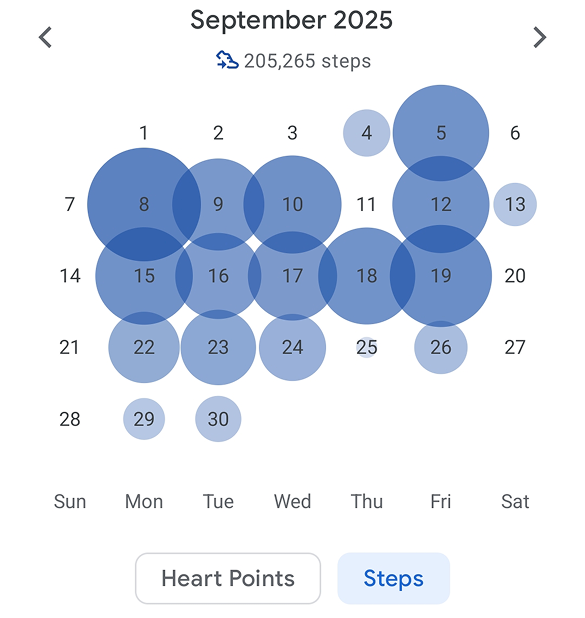}
    \caption{A monthly calendar-based view in Google Fit that visualizes daily step counts using circles.}
    \label{fig:monthly_calendar}
\end{figure}

Calendar-based views (e.g., weekly or monthly; Figure~\ref{fig:monthly_calendar}) offer a powerful and intuitive way to organize and interpret health data in relation to daily life, allowing users to connect data with routines and events. By aligning visualizations with familiar temporal constructs such as days, weeks, and months, these views support pattern recognition and contextual reflection. Visualization libraries should therefore offer flexible calendar-based components, along with abstractions for aggregating, summarizing, and annotating data across temporal scales, to facilitate more meaningful and context-aware health insights.

\subsection{DC5. Support for Smaller \revadd{and Variable} Screens}
% [R3-7] Addresses aspect ratios, orientation, widgets, smartwatches, and foldables under responsive support across mobile form factors.
Beyond primary smartphone applications, mobile widgets and smartwatches introduce both further constraints and opportunities for mHealth visualization libraries, requiring representations optimized for glanceability, efficiency, and seamless integration~\cite{islam2026visualizing}.

\begin{figure*}[t]
    \begin{tabular}{ccp{0.4cm}cc}
        \includegraphics[height=3.4cm]{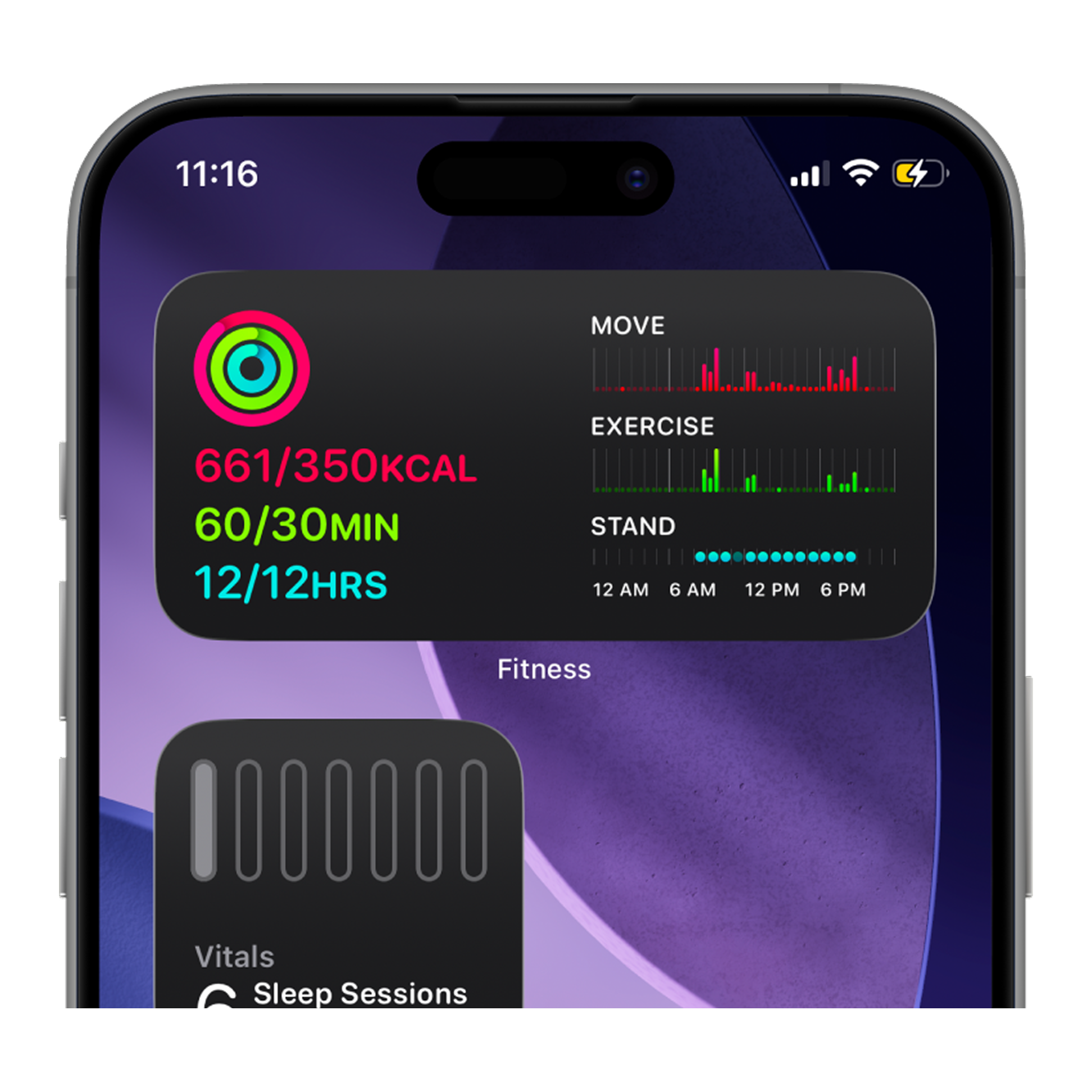} &
        \includegraphics[height=3.4cm]{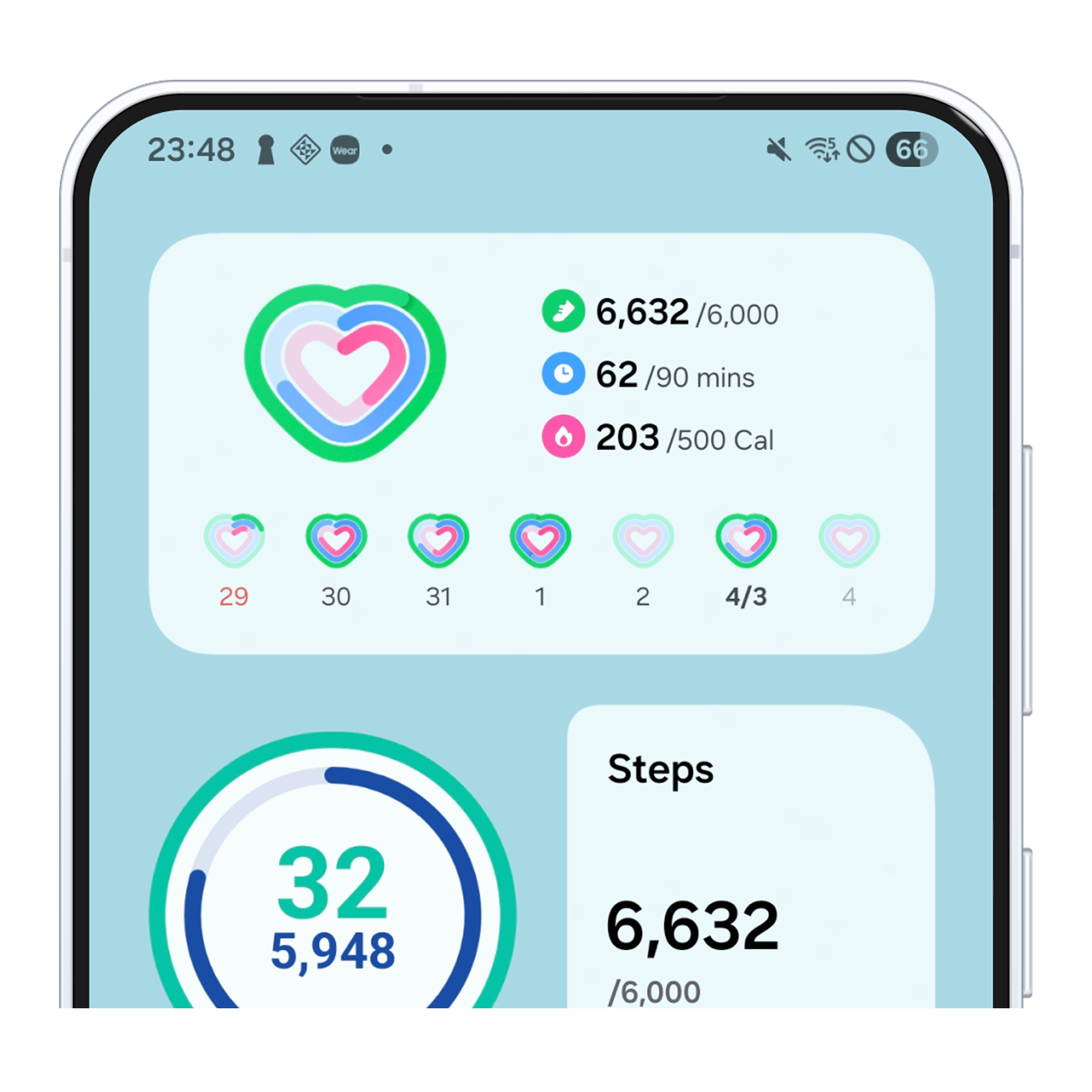} & &
        \includegraphics[height=3.3cm]{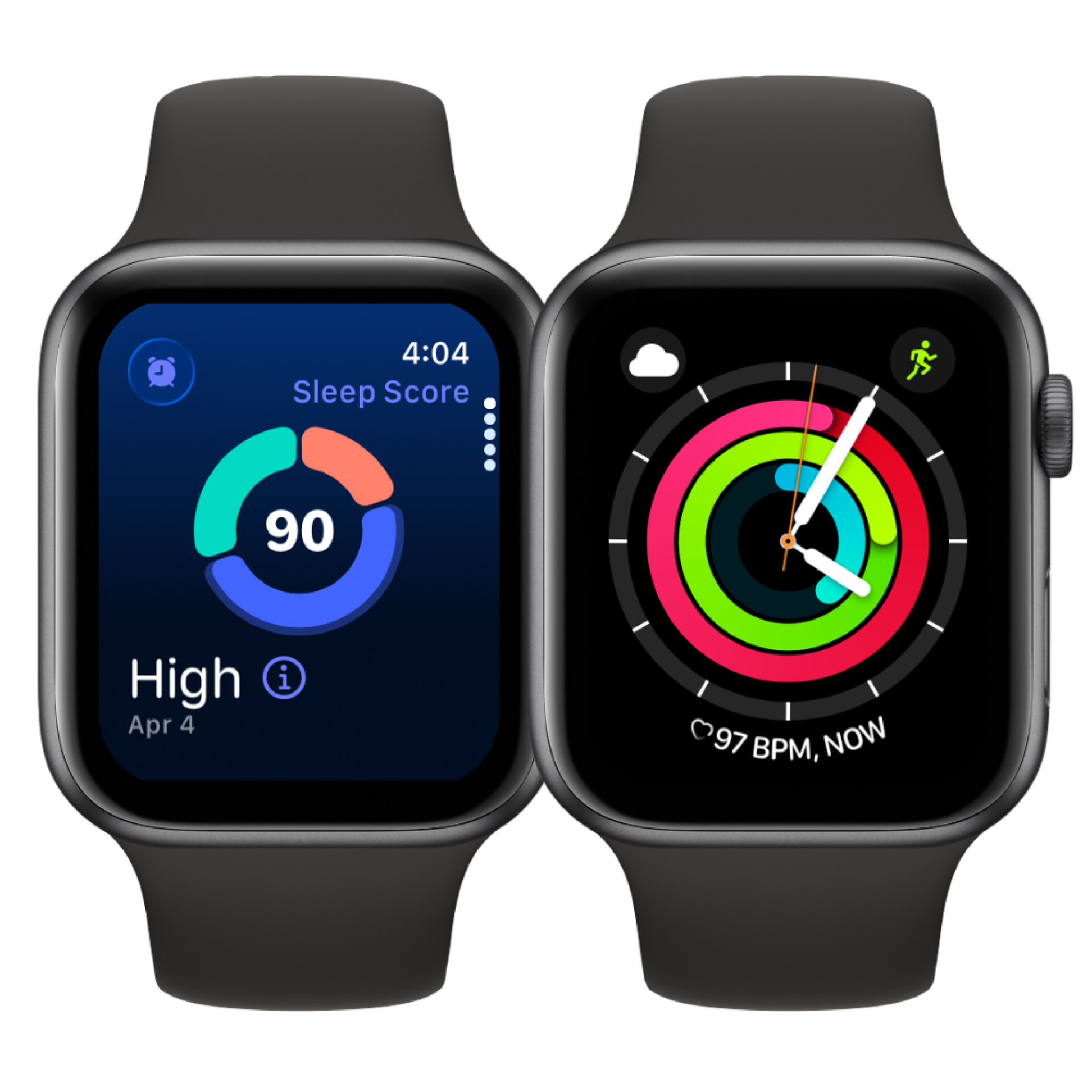} &
        \includegraphics[height=3.3cm]{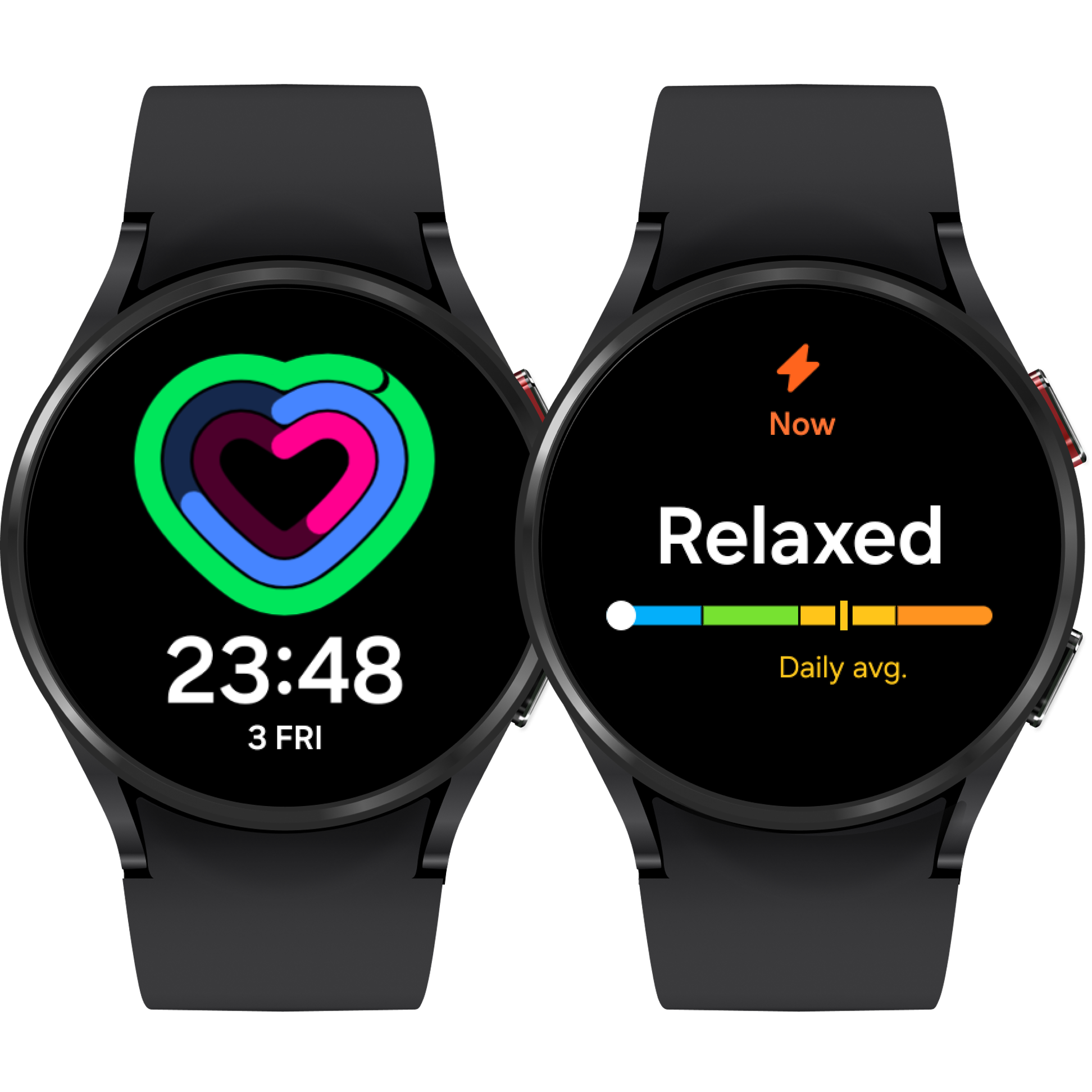} \\
        {\footnotesize Apple Widgets} & {\footnotesize Galaxy Widgets} & & {\footnotesize Apple Watch} & {\footnotesize Galaxy Watch}
    \end{tabular}
    \caption{Example visualizations of health data (e.g., move, exercise, step count, sleep score, and stress) on mobile home screen widgets (left) and smartwatches (right).}
    \label{fig:smallerscreens}
\end{figure*}

Mobile widgets provide glanceable access to personal health data without requiring users to open applications, facilitating continuous, low-friction engagement with health data (e.g., Figure~\ref{fig:smallerscreens}-left). Unlike in-app visualizations, widgets operate within strict spatial, interactional, and system-imposed update constraints, limiting visual complexity and interaction. Their primary role is to support rapid perception, requiring ultra-compact summaries such as minimalist status signals and goal progress that are readily interpretable. %To provide built-in support for creating health data widgets, visualization libraries should support adaptive layouts across widget sizes and configurations while ensuring efficient rendering. 

Smartwatches further extend mobile ecosystem as both continuous sensing platforms and glanceable interfaces\revdel{ for personal data}. While users can interpret simple visualizations on these devices (e.g., Figure~\ref{fig:smallerscreens}-right) \revadd{quickly}\revdel{within hundreds of milliseconds [14]}, their form factor and usage contexts further challenge, including even smaller screen real estate and fleeting interaction windows. \revdel{Furthermore, watch faces necessitate that data visualizations coexist with time and complications without inducing visual clutter.}\revdel{ To bridge the gap between smartphones and wearables, visualization libraries must prioritize highly compact, glanceable representations that emphasize semantic clarity over visual complexity, ensuring flexible integration across diverse watch face layouts.}

\revadd{mHealth visualizations need to remain interpretable not only across varying screen sizes, but also across different aspect ratios and orientations, including on foldable devices. As discussed in DC2, mHealth visualizations commonly involve annotations to convey health-specific semantics such as goals, ranges, and temporal context. Libraries should ensure that visualizations, including these annotations, are responsive to different display contexts.}

\subsection{DC6. Inclusivity and Accessibility by Design}
mHealth visualizations are used by highly diverse populations, including older adults who often have greater and more sustained health management needs, as well as individuals with varying levels of data literacy, health literacy, and prior experience with digital technologies. Users may also have differing sensory, cognitive, or motor abilities that affect how they perceive and interact with visualizations. Designing libraries with these differences in mind is essential to avoid excluding users who could benefit from access to their personal health data.

This requires support for accessible color palettes, readable text and touch targets, adaptable interaction complexity, and alternative representations when appropriate. In some cases, alternative or redundant representations such as simplified summaries, textual explanations, or multimodal cues may be necessary to ensure comprehension. Embedding inclusivity into the library design itself would enable developers to more readily create mHealth apps that are accessible, usable, and meaningful for a broad range of users.

% [R3-8] Adds configurable disclosure for sensitive health data in glanceable and semi-public contexts.
\subsection{\revadd{DC7. Configurable Disclosure of Sensitive Health Data}}
\revadd{Personal health data can be viewed in semi-public contexts, such as lock-screen and home-screen widgets, and notifications. 
These contexts create different disclosure risks because information that is appropriate in the full application, which is typically intended for private viewing, may be too revealing when displayed on a lock screen.} %Although apps and platforms ultimately determine privacy settings, libraries can make privacy-aware display choices easier to implement. Disclosure needs may also vary across situations, because a value suitable in the full application may be too revealing on a lock screen.}

\revadd{Although apps and platforms ultimately determine privacy settings, libraries can make privacy-aware display choices easier to implement. Libraries could expose display modes and defaults that map contexts to appropriate levels of detail, allowing developers to switch among exact values, ranges, trend directions, masked summaries, and abstract status indicators. For example, a glucose widget might show whether a value is within or outside a target range or just whether the value is increasing or decreasing, without revealing the exact reading, while detailed values remain available in the full application. Such support could provide reusable mechanisms for reducing unintended disclosure of sensitive health data.}
\section{DISCUSSION}

We note that the considerations outlined in the previous section are neither exhaustive nor definitive, but instead aim to surface key design directions and open questions.
In this section, we build on these considerations to reflect on the broader implications and tensions involved in developing mobile visualization libraries tailored for personal health data.

\subsection{Static Visualizations as a Starting Point}
Our consideration has focused on static visualizations with annotations as they are well suited for routine monitoring and simple comparison. At the same time, we view this as an initial step. Given the inherently temporal and evolving nature of health data, animation offers a natural and promising extension. Incorporating animation into mHealth visualization libraries could enable richer representations of change over time, guide users' attention, and support smooth transitions between different views or levels of detail, helping users maintain context. Ultimately, animation can enhance users' understanding and engagement with their data.

\subsection{Balancing Generality and Domain Specificity}
A key design challenge in developing mHealth visualization libraries is balancing functional generality with domain specificity. Template-based approaches, tailored to specific health data types (e.g., glucose levels or blood pressure), are a plausible and valuable option: they lower development barriers and embed domain knowledge such as clinically meaningful thresholds and familiar representations. However, predefined templates may over-constrain developers and limit opportunities for flexible and creative use. Instead, we advocate for a more flexible architecture that leverages intelligent defaults while allowing customization, enabling developers to support diverse use cases without sacrificing usability.

% [R1-4] Explains that LLM-powered coding agents can accelerate implementation but do not replace health-specific abstractions, defaults, and semantic components.
\subsection{\revadd{Libraries for AI-Assisted Development}}
\revadd{LLM-powered coding agents can lower the barrier to implementing mobile charts by generating boilerplate code, adapting examples, and accelerating prototyping. However, code generation alone does not eliminate the need for health-specific visualization abstractions, defaults, and semantic components. Without such reusable structures, generated charts may still omit essential features such as target ranges, personal baselines, missing-data indicators, contextual annotations, or configurable disclosure controls. Dedicated mHealth visualization libraries can make these health-specific abstractions available to both human developers and coding agents, enabling the creation of visualizations that are more consistent, interpretable, and easier to validate.}

\subsection{Beyond Health: Generalizing to Other Personal Data Domains}
Although our considerations are motivated by the unique demands of health data, they do not limit generalizability to other domains. Health data impose demanding visualization requirements, including longitudinal structure, personalized baselines, semantic thresholds, and diverse user needs and literacy levels. Designing libraries to address these challenges requires robust abstractions, such as temporal context, semantic annotations, and fluid interaction, that can be extended to other forms of personal data. Many domains, including finance, education, and productivity, share similar characteristics, such as goal-oriented interpretation, temporal variability, and the need for accessible representations for lay individuals. By addressing the complexity and sensitivity of health data, health-data-oriented visualization libraries establish principled foundations applicable to a wide range of personal data-driven mobile applications.

\section{CONCLUSION}
Dedicated mobile data visualization libraries have the potential to accelerate the development of effective mHealth applications while promoting more consistent and interpretable visualization practices. Despite the rapid growth of mHealth technologies, such support remains limited, making this an overdue direction. By embedding visualization expertise directly into reusable libraries, such an approach can raise the baseline quality of health data representations and reduce the implementation burden. Ultimately, developing dedicated mobile data visualization libraries for health data is not merely a technical endeavor, but a design-driven approach to making personal health data more comprehensible and inclusive for diverse users.

\section{ACKNOWLEDGMENTS}
\revadd{We thank the reviewers for their thoughtful feedback and valuable suggestions.
This work was supported in part by the Institute of Information and Communications Technology Planning and Evaluation (IITP) Grant funded by the Korean Government (MSIT), Artificial Intelligence Graduate School Program, Yonsei University, under Grant RS-2020-II201361, and} the Yonsei University Research Fund of 2026-22-0147.
%The Acknowledgments is always plural even if there is a single acknowledgment. The author(s) would like to thank A, B, and C. This work was supported by XYZ under Grant \#\#\#.

%The ``Acknowledgments'' (spelled with just two e's, per American English) section appears immediately after the conclusion and before the reference list. Sponsor and financial support  are included in the acknowledgments section. For example: ``This work was supported in part by the U.S. Department of Commerce under Grant BS123456.'' If support for a specific author is given, then use the following example for correct  wording. ``The work of First A. Author was supported by the U.S. Department of Commerce under Grant BS123456''. Researchers that contributed information or assistance to the article should also be acknowledged in this section, and expressions should be simple and expressed as ``We thank$\ldots$,'' rather than indicating which of the authors is doing the thanking. Also, if corresponding authorship is noted in the paper, it should be placed in the bio of the corresponding author.

\bibliographystyle{IEEEtran}
\bibliography{references}

%\def\refname{REFERENCES}

%\begin{thebibliography}{1}

%\bibitem{AA1}
%G. M. Amdahl, G. A. Blaauw, and F. P. Brooks, ``Architecture of the IBM System/360,'' {\it IBM J. Res. Dev}., vol. 8, no. 2, pp. 87--101, 1964. (journal)

%\bibitem{II1}
%C. J. Smith and J. S. Smith, Rocky Mountain Research Laboratories, Boulder, CO, USA, private communication, 1992. (Private communication)
%\end{thebibliography}\vspace*{-8pt}

%All biographies are limited to one paragraph, following the structure given here: each author's current role and institution (to match the first page of the article); three to  five current research interests; highest degree, topic, and awarding institution (do not include year); professional memberships, such as the IEEE Computer Society and any grade information; and contact information in the form of an email address.

\begin{IEEEbiography}{Bongshin Lee} is a Professor in the Department of Computer Science and Engineering and Vice Director for Research of the Yonsei Institute for Digital Health at Yonsei University, Seoul, Korea. Her research interests include human-data interaction, human-computer interaction, and inclusive data experiences. She received her Ph.D. in Computer Science from the University of Maryland, College Park, MD, USA. Contact her at b.lee@yonsei.ac.kr.
\end{IEEEbiography}

\begin{IEEEbiography}{Seongjae Bae} is a Master's student in the Department of Computer Science and Engineering at Yonsei University, Seoul, Korea. His research interests include human-computer interaction, data visualization, and personal data tracking. Contact him at jae7151@yonsei.ac.kr.
\end{IEEEbiography}

\begin{IEEEbiography}{Mengying Li} is a Ph.D. student in the College of Information at the University of Maryland, College Park, MD, USA. Her current research interests include human-computer interaction, health decision-making, and aging. Contact her at myl99629@umd.edu.
\end{IEEEbiography}

\begin{IEEEbiography}{Eun Kyoung Choe} is an Associate Professor in the College of Information at the University of Maryland, College Park, MD, USA. Her research interests include human-computer interaction, health informatics, and ubiquitous computing. She received her Ph.D. in Information Science from the University of Washington. Contact her at choe@umd.edu.
\end{IEEEbiography}

\end{document}